\documentclass[prd,aps,preprint,amsmath,nofootinbib,amssymb,eqsecnum,showkeys,tightenlines]{revtex4-1}
\usepackage{slashed}
\usepackage{epsfig,latexsym,cancel,amssymb,amsmath,verbatim,mathrsfs}
\usepackage{color}
\usepackage{graphicx}

\def\ra{\rightarrow}
\def\L{\left(}
\def\R{\right)}
\def\wt{\widetilde}
\def\Ld{\Lambda}

\def\f{\frac}
\newcommand{\be}{\begin{equation}}
\newcommand{\ee}{\end{equation}}
\newcommand{\bea}{\begin{eqnarray}}
\newcommand{\eea}{\end{eqnarray}}
\newcommand{\ba}{\begin{array}}
\newcommand{\ea}{\end{array}}

\long\def\symbolfootnote[#1]#2{\begingroup%
\def\thefootnote{\fnsymbol{footnote}}\footnote[#1]{#2}\endgroup}

\newcommand{\beq}{\begin{equation}}
\newcommand{\eeq}{\end{equation}}
\begin{document}

\title{ Post-recombination Dark Matter for the 21-cm Signal}
%Twin-peak Gravitational Waves from Two-step Electroweak Phase Transition}

%\title{Nonradiative EWSB Comes to the Rescue of Gauge Mediated SUSY Breaking: \\
% $\mu/B_\mu$ \& $A_t/m_{H_u}^2-$problem and $(g-2)_\mu$-puzzle}

\author{Zhaofeng Kang}
\email[E-mail: ]{zhaofengkang@gmail.com}
\affiliation{School of physics, Huazhong University of Science and Technology, Wuhan 430074, China}

\date{\today}

%\author{Tao Liu}

%\affiliation{Key Laboratory of Frontiers in Theoretical Physics,
 %            Institute of Theoretical Physics,  Chinese Academy of Sciences,
  %           Beijing 100190,  P. R. China }

\date{\today}

\begin{abstract}

We have no certain knowledge of the early history of dark matter (DM). In this paper we propose a scenario where DM is produced post-recombination but prior to the cosmic dawn. It helps to relax the bounds on DM interactions, in particular with baryons, from the CMB. It may be of interest in some circumstances, for example, to understand the recent cosmic dawn 21-cm signal anomaly. We argue that the cosmic gas cooling mechanism via the minicharged DM-baryon scattering may be viable even if it takes up the total DM budget. We also investigate the possibility of a gluon-philic mediator of a few 10 keV, to find that the most reliable exclusion is from the neutron scattering.

%It may have deep implications to structure formation.

\end{abstract}

\pacs{12.60.Jv,  14.70.Pw,  95.35.+d}

\maketitle

%\section{DM-baryon cooling cosmic gas: no-go versus go}

\section{Introduction and motivation}

%Despite of the lack of convincing particle evidences, dark matter (DM) physics has became an interdiscipline of particle physics, cosmology and astronomy.
Tracking the particle hints of dark matter lies one of the focus of the modern fundamental physics. From the side of particle physics, there are a large pool of models. Among them, the most popular candidate, the weakly interacting massive particle (WIMP), is under siege by many direct and indirect detection experiments~\cite{bounds}. Largely speaking, it has been expelled from the bulk region; for instance, the light WIMP DM not heavier than a few GeV still survives.

The thermal history of dark matter is unknown. The conventional picture for WIMP DM is assuming that it enters the hot plasma at some high temperature, and then decouples/freezes-out before BBN for the sake of not bothering the successful BBN predictions. But kinetic connection with the plasma may still affect the recombination thus CMB, of which we now have good knowledge. The universe falls into the dark ages after recombination. The seed of density fluctuation encoded in DM, that, thanks to the early decoupling, is not damped away by the baryon-photon plasma and leads to the structure formation. The first stars start to form since the cosmic dawn, when is characterized by the coldest cosmic gas $T_b\sim 7$ K, namely the baryon matters having lowest velocity $v_{cd}\sim 10^{-5}$. The information of the cosmic dawn may be seen from the 21-centimetre (cm) signal, due to the atomic hydrogens transitions. For instance, it may reveal interactions between baryons and DM. Considering $v_{cd}\sim 10^{-5}$, it is not surprising that the first light on hydrogen-DM scattering with velocity-dependent may come from the cosmic dawn. One may write the scattering cross section as $\bar\sigma(v)=\sigma_c v_{rel}^{-n}$ with $v_{rel}$ the relative baryon-dark matter velocity.

But usually such information can also be seen from the precise CMB map, because it is likely that the baryon-DM scattering also affects the recombination process. Actually, this consideration strongly limits the window of baryon-DM interactions opening for the 21-cm observations~\cite{McDermott:2010pa,Dvorkin:2013cea,Xu:2018efh}. For 1 GeV DM, the upper bound on $\sigma_c\lesssim 1.1\times 10^{-43}\rm cm^2$ is obtained using the high-$\ell$ polarization data from Planck 2015 release~\cite{Xu:2018efh}.
%,; as DM mass becomes lighter by a factor $r<1$, the upper bound is strengthened roughly by a factor $r^{1/4}$ until the MeV region.

However, an alternative DM history may change such a situation. The recombination happens around $t_R\approx 3.8\times 10^5$ years, while the cosmic dawn starts at around the redshift $z\sim 17$, corresponding to a time scale $t_{cd}\sim 1.8\times 10^8$ years. There is a wide time scale gap between the two eras. So, what if the DM species is just present within this gap? Immediately, the CMB constraint on baryon-DM scattering is removed due to the absence of appreciate DM population during the recombination. The later emergence of DM, following the popular way, could be the heritage from some long-lived decaying mother particle $Y$, whose lifetime $\tau_Y$ is just within the region $(t_R, t_{cd})$. But we should arrange a suppression of DM number density prior to the recombination. Several ways may work: 
\begin{itemize}
\item DM is not never in thermal equilibrium with the visible sector. However, to have the fast baryon-DM scattering, this is cannot be satisfied here.
  \item DM has sizable interactions with the visible sector, so it enters the thermal bath in the early universe. Moreover, DM has effective annihilation channels and thus it can freeze-out with very little relic density.
  \item The reheating temperature is sufficiently low, even below the DM mass. But probably this trick does not work for very low mass DM.
\end{itemize}
So, we will focus on the second option.

Despite of the potential rich applications and implications to dark matter physics,~\footnote{For instance, in this scenario DM may have a large velocity if the mother particle is much heavier than DM, and it may affect the structure formation.} here we demonstrate this scenario in a concrete circumstance: It may be helpful to understand the 21-cm signal anomaly reported by the Experiment to Detect the Global Epoch of reionization Signature (EDGES) Collaboration~\cite{EDGS}. The global 21-cm spectrum reveals a stronger absorption than the maximum predicted by existing models, at a confidence level of 3.8 standard deviations. To explain the anomalously strong absorption, a mechanism is needed to cool the cosmic gas. Dark matter, which decouples at the much earlier time and thus is much cooler than the baryonic gas, can do this job if it elastic scatters with the cosmic hydrogens (either with the electron or nucleons at the microscopic level)~\cite{Barkana}. However, even if the baryon-DM scattering has the largest velocity enhancement $v_{rel}^{-4}$, it is shown by several groups that only the minicharged DM with a small fraction of the total DM budget, $\sim {\cal O}(1\%)$, could explain the anomaly~\cite{Munoz:2018pzp,Berlin:2018sjs,Barkana:2018qrx,Fraser:2018acy}.

The main goal of this paper is to argue that the minicharged DM with 100\% DM fraction may explain the data after relaxing the CMB bound. Additionally, we investigate the possibility of a light gluon-philic mediator.

\section{Minichareged DM}

We now follow the usual argument why the minicharged DM is needed. To enjoy the maximal low-velocity enhancement, one is led to consider the Coulomb-like baryon-DM scattering. Therefore, the mediator $\phi$ mass should be very light,
\begin{eqnarray} \label{Lmd}
 m_\phi\ll m_{\rm DM} v_{cd}=10 {\rm keV}\times(m_{\rm DM}/1{\rm GeV})(v_{cd}/10^{-5}c).
 \end{eqnarray}
This imposes a big challenge viewing from a variety of cosmological/astrophysical constraints on the light particles. Besides, it is nontrivial in model building to accommodate that light particle at the low energy world. Actually, such a light $\phi$ basically negates the possibility of the most popular mediators, such as the keV scale dark photon, for which the kinematic mixing parameter $\epsilon$ has been excluded to below ${\cal O}(10^{-13})$ by the stellar evolution~\cite{H. An}, whereas the even much lighter mediator is even more severely constrained by the fifth force searches. Consequently, the SM photon is the only viable mediator.

Thus let us focus on the minicharged DM here. Such kind of DM may or may not have a gauge origin. The latter can be realized in the 5D setup where the SM hypercharge $U(1)_Y$ in 4D is the zero mode of the bulk $U(1)_Y$~\cite{Mini}. The former can be realized in 4D via the Holdom's approach~\cite{Holdom}:
\begin{eqnarray} \label{}
{\cal L}_{kin}=\bar\chi\L i\gamma^\mu D_\mu -m_\chi\R\chi- \f{1}{4}X^{\mu\nu} X_{\mu\nu} -\f{\kappa}{2} X^{\mu\nu} F_{\mu\nu} -\f{1}{4}F^{\mu\nu} F_{\mu\nu},
 \end{eqnarray}
where $X_{\mu\nu}$ is the field strength tensor of the hidden gauge group $U(1)_X$ with gauge coupling $g_X$, which is in the Coulomb phase; the dark matter candidate $\chi$ is a massive Dirac fermion carrying unit charge under $U(1)_X$, and $D_\mu=\partial_\mu-ig_X X_\mu$. The dark photon $X_\mu$ has kinematic mixing with the SM photon $A_\mu$ via the $\epsilon$-term. One can eliminate the kinetic mixing term through the redefinition of the dark photon field $X_\mu\ra X_\mu-\kappa A_\mu$, and then the kinetic part becomes canonical. The consequence of such an operation is that $X_\mu$ completely decouples from the SM sector whereas $\chi$ obtains a minicharge $\epsilon\equiv g_X\kappa/e$.

The DM in this model has two long range forces, mediated by the dark photon and visible photon. They respectively generate DM-DM self-interaction and DM-baryon/electron interactions. The resulting DM-DM/baryon scatterings are Rutherford scattering, showing the $1/v_{rel}^4$ enhancement at low velocity; $v_{rel}$ is the relative velocity between DM-DM/baryon. For example, the differential cross section of elastic DM-proton scattering is given by
\begin{eqnarray} \label{}
\f{d\sigma_{\rm bDM}}{d\cos\theta}=\f{\epsilon^2e^4}{4m_\chi m_p}\f{2\mu }{(2\pi )4(m_\chi+m_p)}\f{4m_\chi^2m_p^2}{\left[-2\mu^2 v_{rel}^2(1-\cos\theta)- m_\phi^2\right]^2}
 \end{eqnarray}
with $\mu=m_\chi m_p/(m_\chi+ m_p)$ the DM-proton reduced mass and $m_\phi$ the fictitious mass for the photon, which
is added for later convenience. $\theta$ is the scattering angle. The cross section has a divergency related to the forward scattering $\theta\ra 0$, but it is cut at the minimal scattering angle $\theta^*$ due to the Debye screening effects in the plasma~\cite{McDermott:2010pa}. The momentum-transfer cross section~\cite{Dvorkin:2013cea} is
\begin{eqnarray} \label{cross}
\bar\sigma(v_{rel})=\int d\cos\theta (1-\cos\theta)\f{d\sigma_{\rm bDM}}{d\cos\theta}=\f{\epsilon^2e^4}{16\pi}\f{1}{\mu^2 v_{rel}^4}f(\epsilon_\phi^2,\theta^*)
 \end{eqnarray}
where $\epsilon^2_\phi\equiv m_\phi^2/2\mu^2v_{rel}^2$. The function $f$ is defined as
\begin{eqnarray} \label{}
f(\epsilon_\phi,\theta^*)=
\f{\epsilon_\phi^2(\theta^{*2}/2-2)}{(\epsilon_\phi^2+2)
(\epsilon_\phi^2+\theta^{*2}/2)}-
\log\f{\epsilon_\phi^2+\theta^{*2}/2}{\epsilon_\phi^2+2},
 \end{eqnarray}
which in the $\epsilon_\phi\ra 0$ limit numerically is $68-2\log(\epsilon/10^{-6})$~\cite{Munoz:2018pzp}. In the opposite limit one has $f(\epsilon_\phi\gg 1)\ra 4/\epsilon_\phi^4$, corresponding to a heavy mediator.

To explain the 21-cm signal, no parameter space survives if the minicharged DM composes the total DM~\cite{Munoz:2018pzp,Berlin:2018sjs,Barkana:2018qrx,Fraser:2018acy}. However, if the CMB bound is removed, by virtue of the DM production post-recombination, the region $m_\chi\in ({\rm 0.1 GeV, 3GeV})$ with $\epsilon\gtrsim 10^{-6}$ is open; one can see this from the first figure in Fig. 1 of Ref.~\cite{Berlin:2018sjs}. A benchmark point reads $(m_\chi=1{\rm GeV}, \epsilon=2\times 10^{-5})$, and the corresponding cross section is estimated to
\begin{eqnarray} \label{}
\bar\sigma\sim 0.8\times 10^{-18}\L\f{10^{-5}}{v_{rel}}\R^4 \L \f{f(\theta^*)}{70}\R {\rm cm^2},
 \end{eqnarray}
which is around $10^{18}$ pb. It is a huge cross section, compared to the direct detection bounds on the DM-nucleon scattering cross section on the weak scale DM, $\sim 10^{-9}$ pb, but the current technique is not sensitive to DM below the GeV scale.

A similar $\bar \sigma (v_{rel})$ holds for DM-DM scattering after the replacement: $m_p\ra m_{\chi}$ and $\epsilon^{1/2} e\ra g_X$. The strength of DM self-interaction is constrained by the rare Bullet cluster and shapes of dark matter halos of elliptical galaxies/clusters. The latter is found to give the much stronger bound on $(m_\chi, \alpha_X=g_X^2/4\pi)$~\cite{self}. The upper bound on $\alpha_X/m_\chi$ is linear; for instance, $\alpha_X\lesssim 2\times 10^{-7}$ for $m_\chi=1$ GeV. On the other hand, the cross section of annihilation channel $\bar \chi\chi\ra X+X$ is proportional to $\alpha_X^2/m_\chi^2$, and therefore DM cannot sufficiently deplete its relic density by annihilating into the dark radiations. In other words, the dark photon may be irrelevant to the DM phenomenologies of interest. One has many options to deplete the DM number by introducing extra gauge/Yukawa interactions for $\chi$, and we do not expand discussions here.

To realize the late decay production of $\chi$, we consider a SM extension by a complex scalar $S$, with the following relevant terms,
\begin{eqnarray} \label{}
{\cal L}_{S}=\eta_1|S|^2|H|^2+\L\mu_s^2 S^2+\eta_2\f{S^2}{M_P}\bar\chi\chi+c.c.\R,
 \end{eqnarray}
with $M_P$ the Planck scale. This is a working model for our phenomenological purpose. One may understand its structure like this: $S$ is charged under a global $U(1)$, which is softly broken to the $Z_2$ subgroup by the $\mu_s$ term and the dimension-five operator. The singlet is assumed to acquire a VEV and then we decompose $S=(v_s+s+i A)/\sqrt{2}$. Without the dimension-five operator, the CP-odd component $A$ is a stable DM candidate, and it is thermalized via the Higgs portal and freezes-out as usual, leaving a proper relic density $\Omega_{A}h^2\simeq 0.2 m_\chi/m_A$~\cite{CDM}. If kinematically accessed, after decoupling $A$ decays into a pair of $\chi$ via the dimension-five operator, with a lifetime estimated to be
\begin{eqnarray} \label{}
\tau\sim\eta_2^{-2} \L\f{M_P}{v_s}\R^2\f{4\pi}{m_A}\sim 10^{13}s\times\L\f{ 10\rm GeV}{\eta_2 v_s}\R^2 \L\f{1\rm GeV}{m_A}\R.
 \end{eqnarray}
So, a very long-lived $A$ is naturally obtained for $v_s$ around the weak scale.

\section{Is a gluon-philic mediator possible?}

In the absence of the CMB bound, we now move to other possibilities, e.g., those with a fairly light mediator between DM and hydrogen. Although we find that no mediator is allowed confronting with the constraints, it is still meaningful to present our exploration, because a light gluon-philic mediator is relatively new and may be of interest elsewhere.

A dark CP-even/odd Higgs boson does not work in a natural way, if it couples to the SM fermions via mixing with the SM-like Higgs boson or an extra CP-odd Higgs boson, the latter furnished in the simple extension such as 2HDM; the reason is that generically the mixing angle between the keV mediator and the weak scale bosons are supposed to be exptremely small, barring huge fine-tuning to maintain a keV sector confronting the large scale inherited from the mixing portal. Exclusive couplings between light quarks to unavoidably induce the coupling between $\phi$ and photons through the quark loops.

Therefore, we are left with the possibility that the mediator interacts with the nucleons via the gluons, e.g., the well-known dimensional-five operators (for the CP-even and -odd mediator, respectively),
\begin{eqnarray} \label{}
\f{1}{\Ld}\f{\alpha_s}{8\pi} \phi G_{\mu\nu}^aG^{a\mu\nu},\quad \f{1}{\Ld}\f{\alpha_s}{8\pi} \phi G_{\mu\nu}^a\wt G^{a\mu\nu},
 \end{eqnarray}
where $G^{\mu\nu}$ is the gluon field tensor and $\wt G^{\mu\nu}$ its dual. In a complete theory, one has to suppress similar operators with gluon replaced by photon, because they, similar to the bounds on axion-like particles, are strongly constrained by astrophysics. A pure gluonic operator can be obtained via a purely colored loop. The constraints on $\Ld$, to my knowledge, are fairly weak. For the CP-even case, in the absence of mixing the pion, the known constraint comes from the mono-jet search at the LHC assuming a long-lived $\phi$~\cite{Mimasu:2014nea},
\begin{eqnarray} \label{constrain1}
\f{1}{\Ld}\lesssim  0.6\times 10^{-2}\rm GeV^{-1}.
 \end{eqnarray}
For the CP-odd mediator case, the pion-$\phi$ mixing effect results in a mildly tighter (about one order of magnitude) constraint from $K^+\ra \pi^++a$ decay.

The effective theory for $\phi$ and nucleon interaction can be constructed through sandwiching the gluon operators between the nuclear states at rest. For the CP-even gluonic operator one has~\cite{SUDM}
\begin{eqnarray} \label{}
\langle N| GG|N\rangle= \f{16\pi}{9\alpha_s}m_N f_{T_G}^{(N)}\bar NN
 \end{eqnarray}
with $f_{T_G}^{(N)}\approx 0.83$ for $N=n, p$. Whereas for the CP-odd gluonic operator, the nuclear matrix barely gains attention in the new physics community. But recently it has been studied in Ref.~\cite{Bishara:2017pfq} and we borrow their result here
\begin{eqnarray} \label{}
\langle N| G\wt G|N\rangle=-m_N\Delta u \f{8\pi}{\alpha_s}  \bar Ni\gamma^5N
 \end{eqnarray}
with $\Delta \approx 0.9$. Therefore, the resulting effective Lagrangian for $\phi-N$ interactions are given by
\begin{eqnarray} \label{}
{\cal L}_{\phi NN}&=& g_{\phi NN} \phi \bar N N,
\quad
F_{\phi NN}= \f{2}{9}f_{T_G}^{(N)}\f{m_N}{\Ld}, \\
{\cal L}^5_{\phi NN}&=& g^5_{\phi NN} \phi \bar N i\gamma^5 N, \quad
F^5_{\phi NN}= -\Delta_u \f{m_N}{\Ld}.
 \end{eqnarray}
They, along with the $\phi$ and dark matter interactions, furnish the simple effective Lagrangian to describe DM-baryon scattering
\begin{eqnarray} \label{}
{\cal L}_{\phi \chi\chi}&=&g_{\phi \chi \chi} \phi\bar\chi \chi ~~~~{\rm or} ~~~~g^5_{\phi \chi \chi}  \phi\bar\chi i\gamma^5 \chi.
 \end{eqnarray}
Of course, DM can participate other interactions that determine DM relic density we do not specify here.

The CP-odd case seems to be of more theoretical interest because its lightness can be a natural consequence of symmetry breaking. Nevertheless, in the non-relativistic limit the DM-baryon scattering is suppressed by $v_{rel}^4$, which is clear in the heavy dark matter effective theory where $\bar \chi i\gamma^5 \chi\ra -i/m_\chi (\bar \chi_v \vec{q}\cdot \vec{S}_\chi \chi_v)$ with $\vec q$ the transferred three spatial momentum, $\vec S_\chi$ the spin of $\chi$, which ~\cite{Bishara:2017pfq}; the similar limit applies to the nucleons. Hence, in what follows we just consider the CP-even mediator case.

In terms of the effective interactions and Eq.~(\ref{cross}), the momentum-transfer cross section, in the absence of velocity enhancement is calculated to be
\begin{eqnarray} \label{}
\bar\sigma=\f{g_{\phi \chi \chi}^2g_{\phi NN}^2}{\pi}\f{\mu^2}{m_\phi^4}=0.3\times10^{-18}{\rm cm^2}\L\f{g_{\phi \chi \chi}g_{\phi NN}}{2\times10^{-4}}\R^2\L\f{10^{-9}\rm GeV^{2}}{m^2_\phi}\R^2\L\f{\mu}{\rm 0.5 GeV}\R^2,
 \end{eqnarray}
With the above results we turn to the requirements on effective couplings. For $m_\phi\sim {\cal O}(10){\rm keV}$, regardless the stellar bound which is unclear in the strong coupling region~\cite{Hardy:2016kme}, the robust laboratory bounds from the neutron scattering gives an upper bound $g_{\phi NN}\lesssim 10^{-6}$. Therefore, the light mediator scenario fails in explaining the 21-cm signal anomaly.

%DM-baryon scattering leads to DM-DM scattering, again exchanging $\phi$. It is tempting to use it to solve the small scale crisis~\cite{Spergel:1999mh}. But things never come so easy. To solve the small scale problems at the same time evading the exclusions from the larger scale systems such as the Bullet cluster, the transition between the contact and Rutherford limits should occur at the velocity $v_t\sim 300 {\rm km/s}$, corresponding to the transition between the dwarf and cluster scales. In other words, the mediator mass $m_\phi \sim m_\chi v_t\sim 10^{-3}m_\chi$~\cite{Tulin:2017ara}, which is much heavier than the upper bound given in Eq.~(\ref{Lmd}). However, it is of interest to conjecture that the CP-even scalar boson $\phi$ is a dark Higgs, escorted by a massive gauge boson $X_\mu$, and they are members of a dark gauged $U(1)_X$ sector to address the full dynamics of dark matter. By the accidental reason, for instance the dark Higgs self-coupling is much lighter than the dark gauge coupling, $X$ turns out to be much heavier than $\phi$, at least by two orders of magnitude.~\footnote{If the $U(1)_X$ spontaneously breaking is by the Coleman-Winberg mechanism~\cite{1}, the mass of dark Higgs boson is loop suppressed compared to the dark gauge boson, and thus their mass hierarchy may be natural; see one example~\cite{Guo:2015lxa}.}

\section{Conclusions}

Recently the first cosmic dawn 21-cm signal is reported, indicating that a (light) dark matter may have a sizable elastic scattering with hydrogen. But the only viable dark matter candidate is found to be a minicharged dark matter with a mini fraction of the total dark matter budget. The necessity of multi-component dark matter is mainly owing to the CMB constraint. It motivates us in this paper to consider the possibility that most of the dark matter is produced after recombination, and the CMB bound is removed
thus reopening the window for the single DM species and even other dark matter candidates such as those with a light mediator. The proposal is of interest to any other DM circumstances that suffer the CMB constraint. It may also have deep implications to structure formation.

\noindent {\bf{Acknowledgements}}
This work is supported in part by the National Science Foundation of China (11775086).

\appendix

%\section{Bound states}

\end{document}